\documentclass{article}
\usepackage{spie}
\usepackage{epsfig}

\title{Removing sky contributions from SCUBA data}

\author{T.\ Jenness\supit{a}, J.\ F.\ Lightfoot\supit{a,b} and
W.\ S.\ Holland\supit{a}
\skiplinehalf
\supit{a}Joint Astronomy Centre, 660 N.\ A`oh\={o}k\={u} Place, University Park, Hilo, HI 96720, USA
\skiplinehalf 
\supit{b}Royal Observatory, Blackford Hill, Edinburgh, EH9 3HJ, United Kingdom
}

\authorinfo{Other author information: (Send correspondence to T.J.)\\T.J.:
E-mail: t.jenness@jach.hawaii.edu\\ J.F.L.: E-mail: j.lightfoot@roe.ac.uk\\
W.S.H.: E-mail: w.s.holland@jach.hawaii.edu}
\begin{document} 

  \maketitle

%%%%%%%%%%%%%%%%%%%%%%%%%%%%%%%%%%%%%%%%%%%%%%%%%%%%%%%%%%%%% 

\begin{abstract}

The Submillimetre Common-User Bolometer Array (SCUBA) is a new continuum
camera operating on the James Clerk Maxwell Telescope (JCMT) on Mauna Kea,
Hawaii. It consists of two arrays of bolometric detectors; a 91 pixel 350/450
micron array and a 37 pixel 750/850 micron array. Both arrays can be used
simultaneously and have a field-of-view of approximately 2.4 arcminutes in
diameter on the sky.

Ideally, performance should be limited solely by the photon noise from the sky 
background at all wavelengths of operation. However, observations at
submillimetre wavelengths are hampered by ``sky-noise'' which is caused by
spatial and temporal fluctuations in the emissivity of the atmosphere above
the telescope. These variations occur in atmospheric cells that are larger
than the array diameter, and so it is expected that the resultant noise will
be correlated across the array and, possibly, at different wavelengths.

In this paper we describe our initial investigations into the presence of
sky-noise for all the SCUBA observing modes, and explain our current technique 
for removing it from the data.

\end{abstract}

%>>>> Please include a list of keywords after the abstract 

\keywords{Submillimetre astronomy, Bolometer arrays, SCUBA, sky-noise}

%%%%%%%%%%%%%%%%%%%%%%%%%%%%%%%%%%%%%%%%%%%%%%%%%%%%%%%%%%%%%

\section{INTRODUCTION}

\label{sect:intro}  % \label{} allows reference to this section

Observations at submillimetre wavelengths have always been severely hindered
by the atmosphere, which is only partially transparent throughout the region.
In addition to attenuating the signal, the atmosphere and the immediate
surroundings of the telescope and the observatory contribute thermal radiation 
which is often a few orders of magnitude greater than the signal from the
source under investigation. The necessity to extract the source signal from
this background has led to the techniques of sky chopping and telescope
nodding. 

The aim of SCUBA is to be background limited, i.e.\ limited by the photon
noise from the sky background.  The main limit to sensitivity, particularly at
the shorter submillimetre wavelengths arises from sky-noise. Sky-noise
manifests itself in three ways: a D.C.\ offset, noise due to sky variability
and scintillation. For this to be achieved the DC offset and effects of sky
variability and scintillation must be removed. Chopping and nodding remove the
DC offset and, in addition, diminish the effect of sky variability but don't
remove it completely.

\begin{figure}
\begin{center}
\epsfig{width=5.0in,file=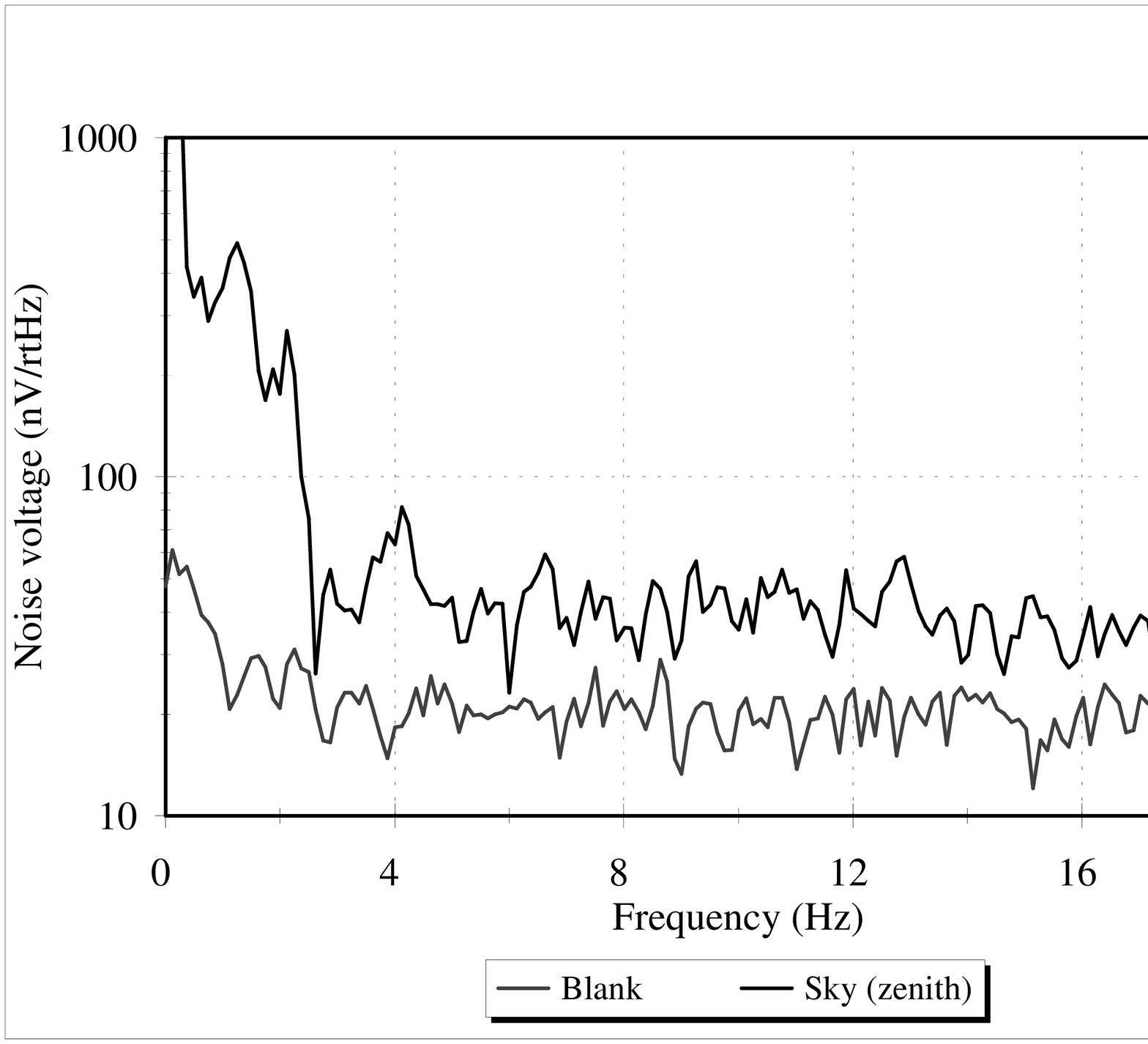,clip=}
\caption{This figure shows noise spectra for a single pixel at 850
microns with no chopping (upper line) along with the system noise (lower
trace). Clearly excess noise is visible below 3~Hz when looking at the sky.}
\label{fig:bolnoise}
\end{center}
\end{figure}

Fig.\ \ref{fig:bolnoise} shows the noise spectrum for the centre pixel of the
long-wave (LW) array at 850~microns. The bottom trace illustrates the system
noise (i.e.\ the intrinsic detector and amplifier noise). The top trace shows
the corresponding spectrum when the detector views the zenith sky under good
observing conditions at 850~microns (approximately 1~mm of precipitable water
vapour)  with the chopper stationary. It can be seen that there is a ``1/f''
component to the zenith sky trace, extending out to 4~Hz or so, which is well 
above the system noise level.

Duncan et al.\cite{DRAC95} have shown that two-position chopping at a
frequency of approximately 8~Hz greatly reduces the sky-noise contamination in
the power spectrum, but that on many occasions there is still a level of noise
well above the system noise level.

There are several ways to reduce the residual sky-noise; firstly careful
design of the filters to select the most transparent parts of the atmospheric
window. Secondly, the chop throw should be kept as small as possible since the 
magnitude of sky-noise increases approximately linearly with chop throw
amplitude\cite{DRAC95}. Three-position chopping has also been found to produce 
considerable improvements. In this paper we concentrate our discussion on
how to remove the error due to sky variability from array data taken with 
the Submillimetre Common-User Bolometer Array (SCUBA)\cite{HGL98,CGD94,GC95}
at the James Clerk Maxwell Telescope (JCMT).

\section{Observing Modes with SCUBA}
\label{sect:obsmodes}

Three different observing modes are available\cite{LDG95}: two imaging
modes (one for sources smaller than the array (jiggle mapping) and one
for sources larger than the array (scan mapping)) and a photometry
mode for measuring the flux density of point sources. The jiggle
mapping and photometry modes have much in common and can be grouped
together as `jiggle modes' (a photometry observation is simply an
under-sampled jiggle map).  All these modes suffer from sky variability to a
varying degree and each mode will be investigated in turn.

\subsection{Jiggle modes}

Both photometry and jiggle maps share the same basic observing method.
The secondary mirror is chopped between the source position and sky at
approximately 8~Hz to remove the bulk of the sky emission from the
signal.  SCUBA takes data at 128~Hz but currently hardware restriction
mean that samples have a fixed integration time of 1
second\footnote{The transputers pre-process the data using a digital
demodulation technique to calculate the chopped signal for 1 second}.
In order to remove slowly varying background emission and telescope
asymmetries the telescope is nodded so that the source appears in the
opposite beam.  The nods are then subtracted before further processing
so that each data point is in fact the difference between two 1 second
integrations taken at different times -- the time difference is
directly related to the number of samples taken between each nod.  It
should also be noted that this technique results in pseudo-three beam
data so that it is assumed that there is no source emission at any of
the off-positions.

\subsubsection{Photometry}

The photometry mode is used to measure the flux density of point
sources.  In general a bolometer is centred on the source and a small
jiggle pattern is used to compensate for any offsets between the two
arrays: the standard jiggle pattern is a 3 by 3 grid with 2 arcsec
spacing. For this configuration it takes $2\times9$ seconds to complete an
integration (nine seconds per nod) plus some overhead for moving the
telescope. One important point is that data are taken for all
bolometers even though only one is on source at any given time.

\subsubsection{Mapping sources smaller than the array}
\label{sect:jigmap}

\begin{figure}
\begin{center}
(a) \epsfig{width=2.35in,file=jig16.eps}
(b) \epsfig{width=3.0in,file=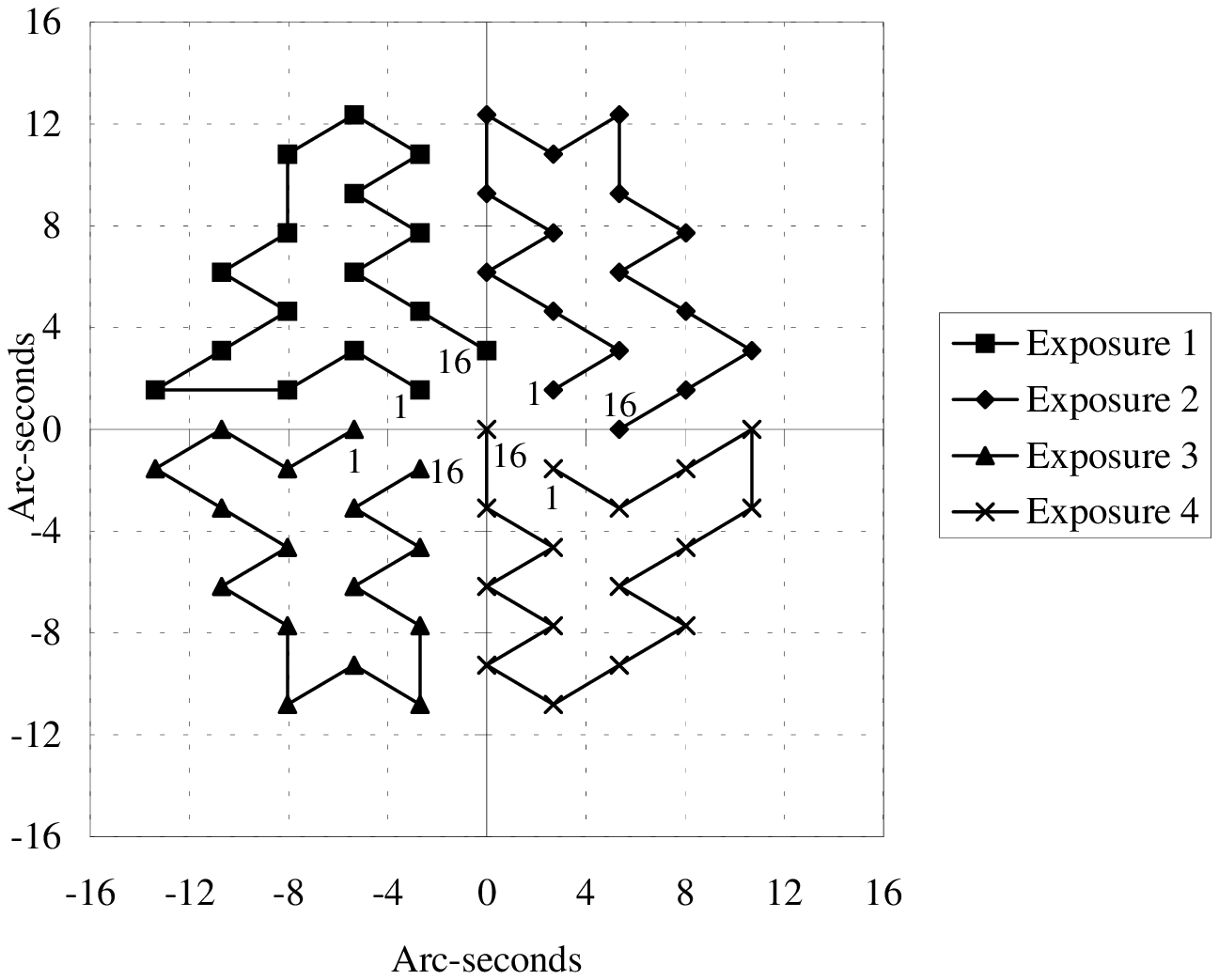,clip=}
\caption{(a) The jiggle pattern used to fully-sample the long-wave
array. The six nearest bolometers are shown to indicate the hexagonal packing
of the array. (b) The jiggle pattern used to fully sample both array. Note
that it is split into four sections so that the telescope can nod four times
during an integration.}
\label{fig:jigpattern}
\end{center}
\end{figure}

As shown in Fig.\ \ref{fig:jigpattern}a the individual bolometers are
arranged in a hexagonal pattern to maximize packing efficiency. The
size of the feedhorns means that the bolometers are separated by
approximately 2 beam-widths so a fully-sampled image can not be taken
in one go. To overcome this problem the secondary mirror is `jiggled' to
fill in the gaps so that a fully-sampled image can be obtained. For a
single array 16 different secondary mirror positions are
required for this but 64 positions are required when both arrays are being used
simultaneously because of the factor of 2 difference in bolometer spacing
between the arrays (see Fig.\ \ref{fig:jigpattern}b).

\subsection{Mapping sources larger than the array}

Jiggle mapping is not useable when there is a possibility of source
emission in the off-beam. In this case a raster mapping technique is
used where the telescope scans across the desired area whilst
chopping.

\section{Investigation of sky variability}

This section will describe the differing ways in which sky variability
manifests itself and the ways it can be removed.

We can only hope to remove sky variability if the size of the sky features are
much larger than the array. The scale height of the `screen' layer in the
atmosphere that is moving over the telescope and causing the sky variation is
of order 2~km or less. Combining this with typical wind speeds and the
observed frequency of the fluctuations implies that the scale size of the
fluctuations is of order 1000 arcsec or more so the error caused on something
the size of the array should be well approximated by a DC offset.

\subsection{Jiggle modes}

The choice of chop configuration is critical for minimising the
observed sky variability.  In most cases azimuth chopping and nodding is
used since this has been shown to be the best configuration for
reducing systematics\cite{CLH93} but occasionally different chop
directions are required (e.g.\ a chop in RA/dec to a fixed position on
the sky or chopping between two bolometers on the array). The other
important factor is the size of the chop throw -- as the chop throw
increases the sky cancellation becomes worse and the beam shape
becomes more distorted.

\subsubsection{Sky removal technique}

The standard data reduction technique for jiggle data is:

\begin{enumerate}
\item The first stage in the data reduction process is to subtract the
    nods. This means that each processed data point is the
    difference of two one second samples taken 9 or 16 seconds apart.

\item Correct for atmospheric extinction and flatfield the data.

\item Identify sky bolometers or sky regions. For extended sources it is
preferable to specify a sky region for all but the shortest observations.
SCUBA is mounted on the nasmyth platform of an alt-az telescope with no image
rotator such that during a long observation bolometers would see sky and
source.

\item For each 2 second sample (2 seconds because of the nodding) calculate
the average sky signal and subtract this from the entire array.

\end{enumerate}

Currently no attempt has been made to remove a sky plane from the data.
Preliminary investigations have been made but in general the effect has been
small relative to the large scale DC offset.

\subsubsection{Photometry}

For most photometry observations the sources are known to be isolated
and point-like and the chop throw can therefore be chosen to minimise
systematics: two position chopping and nodding in azimuth are used\cite{CLH93} 
with small chop throws (30 to 60 arcsec).

\begin{figure}
\begin{center}
\epsfig{width=5.0in,file=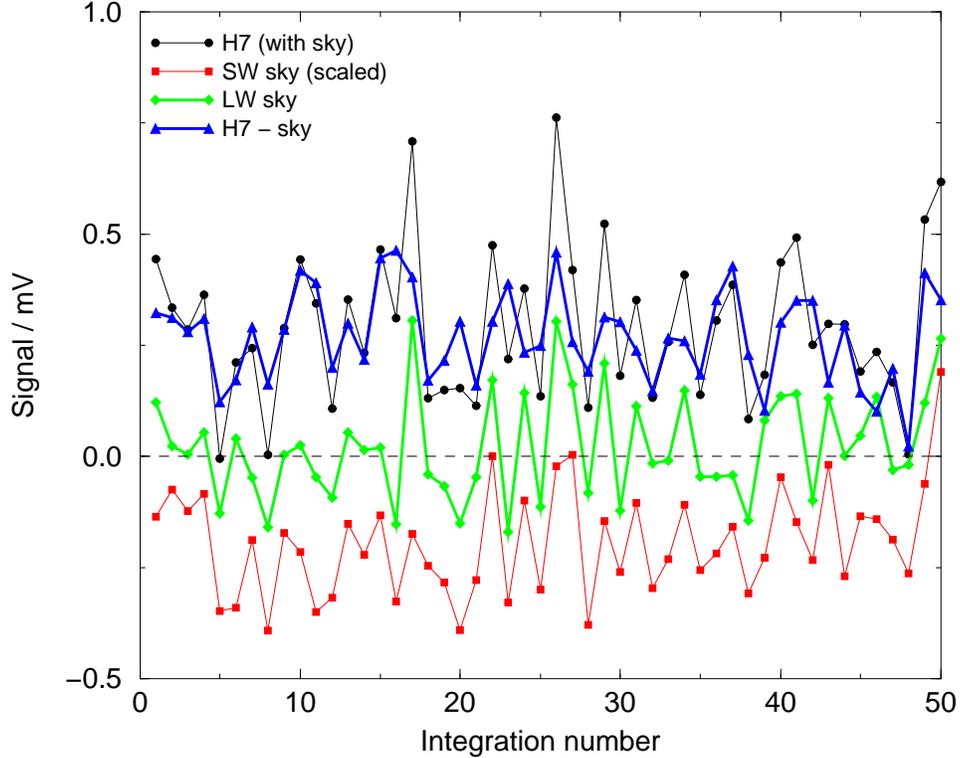,clip=}
\caption{Sky and source signal for a 50 integration photometry
observation at 850-microns. The lower trace shows the sky signal for the
short-wave array (reduced by a factor of 5). In this case sky removal improves 
the signal-to-noise by a factor of 2.}
\label{fig:photex}
\end{center}
\end{figure}

Fig.\ \ref{fig:photex} shows an example of some 850-micron photometry data
with a sky transmission of approximately 80 per cent and seeing of
1.25~arcsec.\footnote{The seeing monitor is operated by the Smithsonian
Astronomical Observatory} For these data the measured sky variability
contribution is approximately 0.12~mV (or approximately 30~mJy/beam) and when
removed boosts the signal-to-noise by a factor of 2.

\begin{figure}
\begin{center}
\epsfig{width=5.0in,file=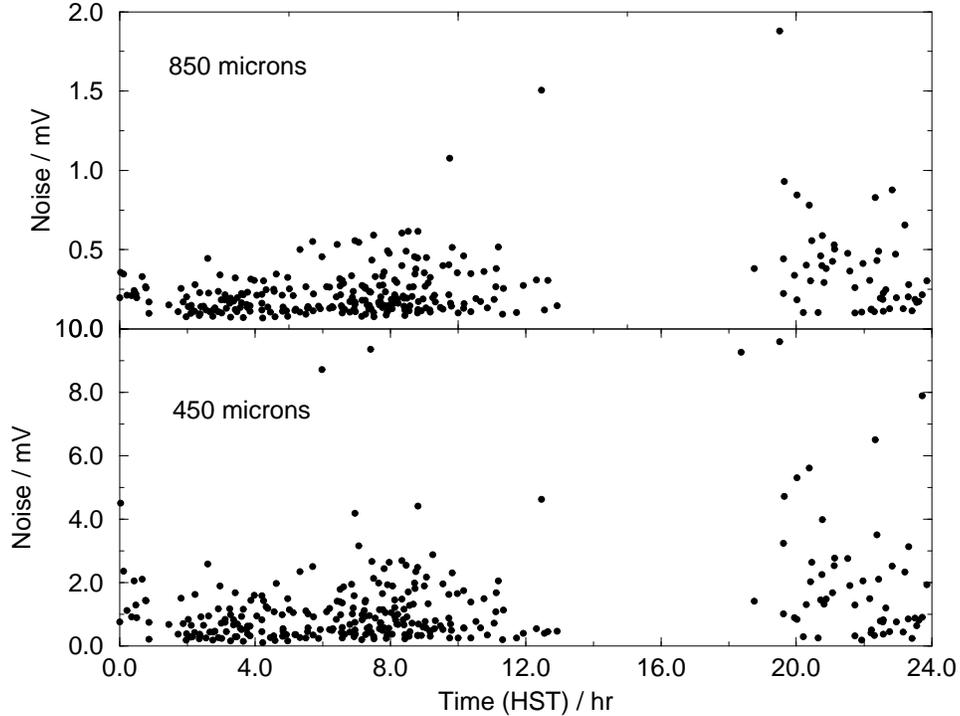}
\caption{Magnitude of the measured sky variability plotted against the time of day.}
\label{fig:timeseries}
\end{center}
\end{figure}

All of the long integration photometry data\footnote{i.e.\ `long' is defined
as observations consisting of at least 40 separate integrations.} taken
between 1997 October and 1998 February have been processed in order to
investigate sky variability for a set of different observations. It was
expected that sky variability would increase during the daylight hours but, as
shown in Fig.\ \ref{fig:timeseries}, these data do not show a large increase
in sky variability during the morning. Although the number of data points
taken before midnight are small there is a suggestion that the sky variability
is higher for the first part of the night.  Whilst these results seem to
contradict the data presented later on for 64 position jiggle maps it must be
remembered that photometry observations have much shorter integration times,
chop throws and nod cycles. Additionally it is also possible that the months
chosen are special in some way and more data analysis is required.

\begin{figure}
\begin{center}
\epsfig{width=5.0in,file=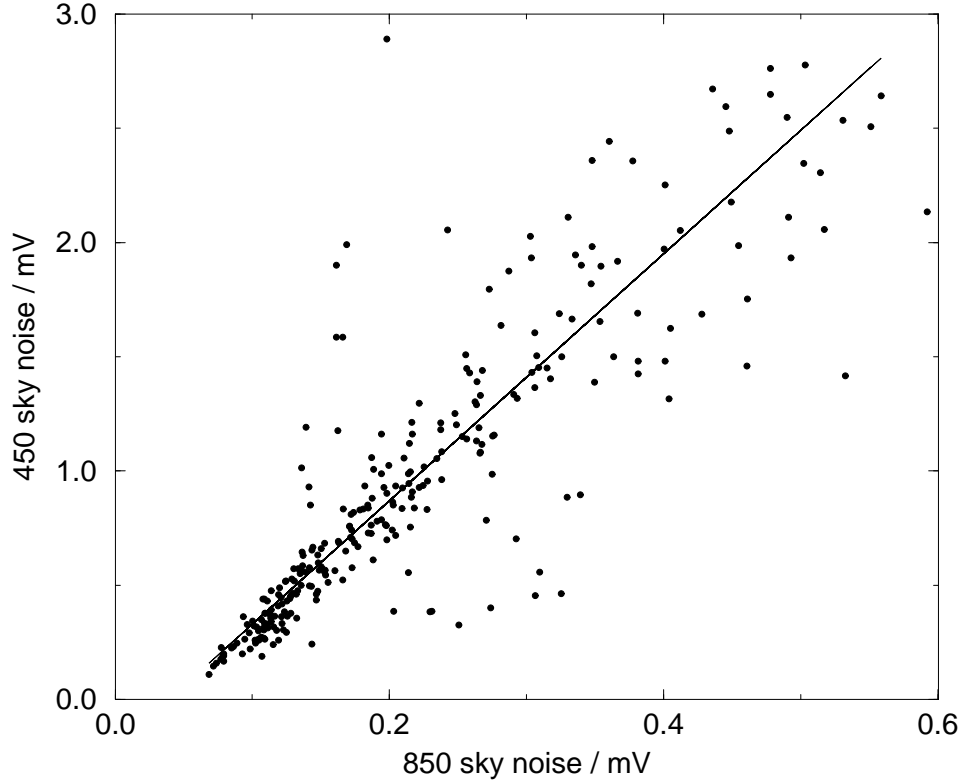}
\caption{Sky variability at 450 microns plotted against the sky variability for the short
wave array for long photometry observations. A linear least squares fit is
shown: the gradient is approximately 5.0 with a correlation coefficient of
more than 0.9.}
\label{fig:ratio}
\end{center}
\end{figure}

Fig.\ \ref{fig:ratio} shows data for 250 observations with a linear least
squares fit. Although the correlation is not perfect this preliminary result
indicates that the magnitude of the sky variability seen on the short-wave
array is proportional to that seen on the long-wave array.

\subsubsection{Mapping}

Jiggle mapping suffers in two ways when compared with photometry mode:
\begin{enumerate}
\item The array has a diameter of more than two arcminutes, so it is
necessary to chop at least this distance to avoid chopping onto the
array.
\item The size of the jiggle pattern dictates that nodding will take
place every 16 seconds. Whilst taking data on both arrays the jiggle
pattern consists of 64 points, but a nod every 64 seconds is far too
long even in the most stable conditions. For this case the pattern is
split into 4 chunks of 16 so that nodding occurs every 16 seconds as
for a single array observation (see Fig.\ \ref{fig:jigpattern}b).
\end{enumerate}

\begin{figure}
\begin{center}
(a)\epsfig{width=3.0in,file=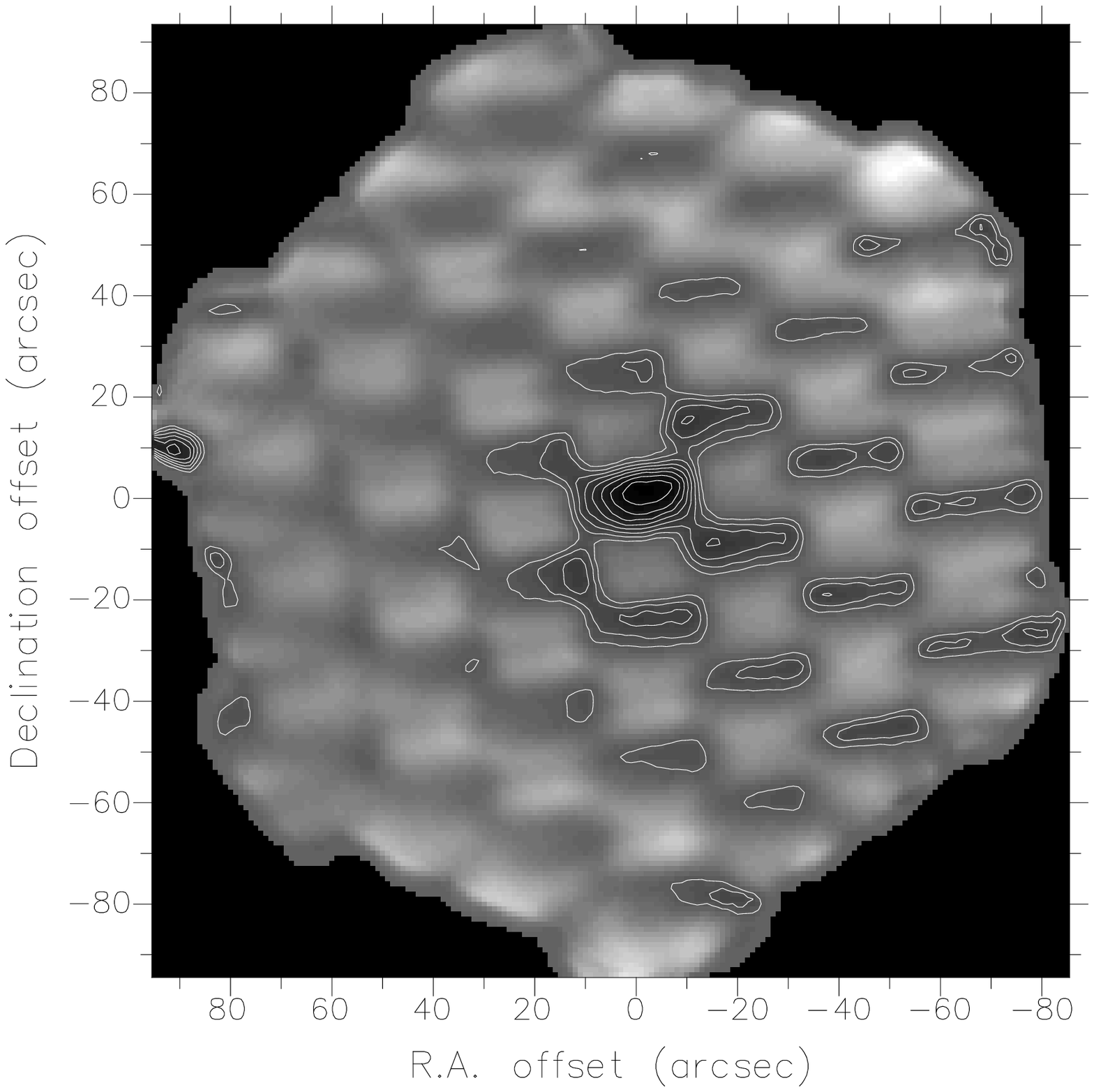,clip=}
(b)\epsfig{width=3.0in,file=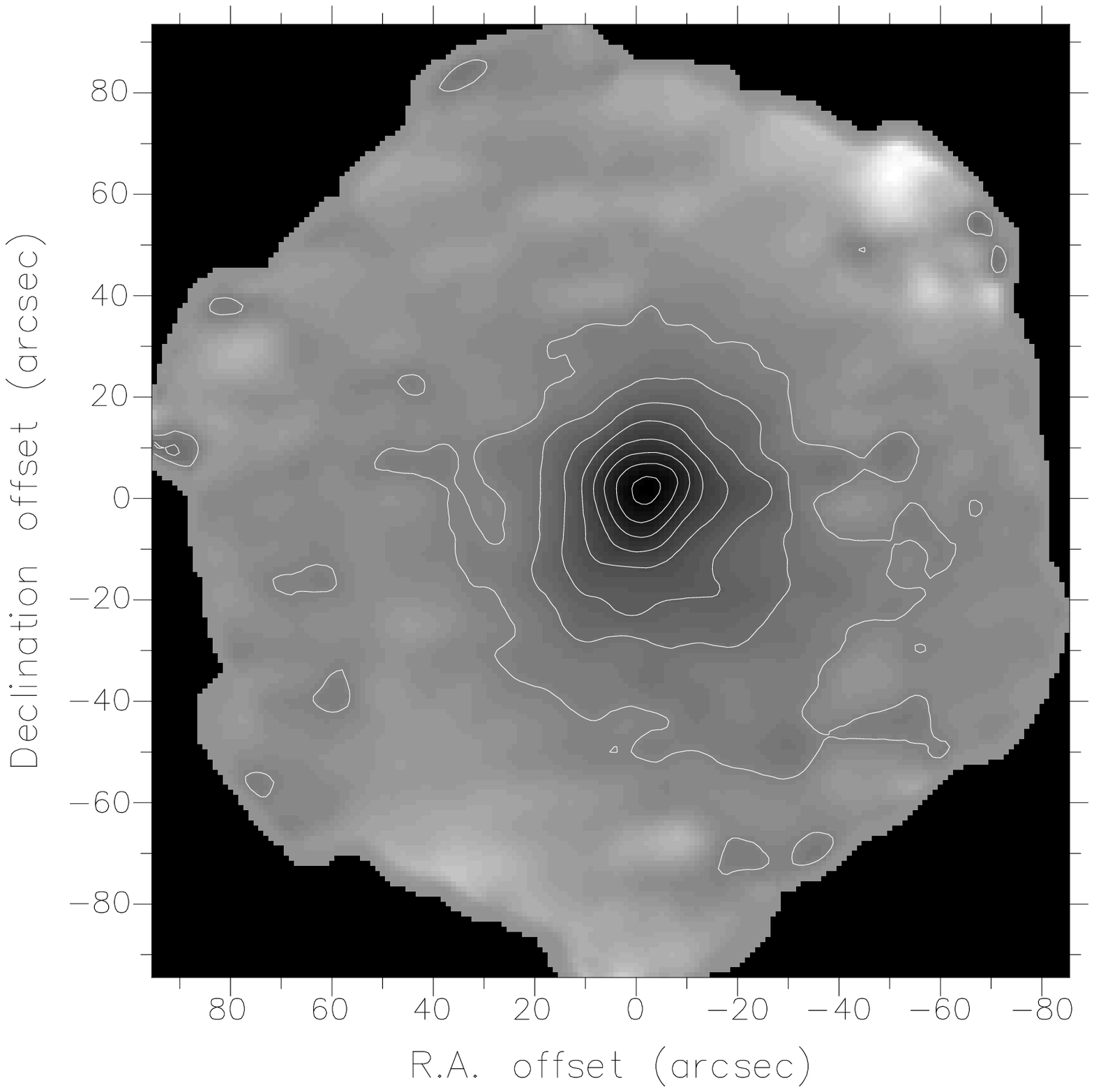,clip=}
(c)\epsfig{width=5.0in,file=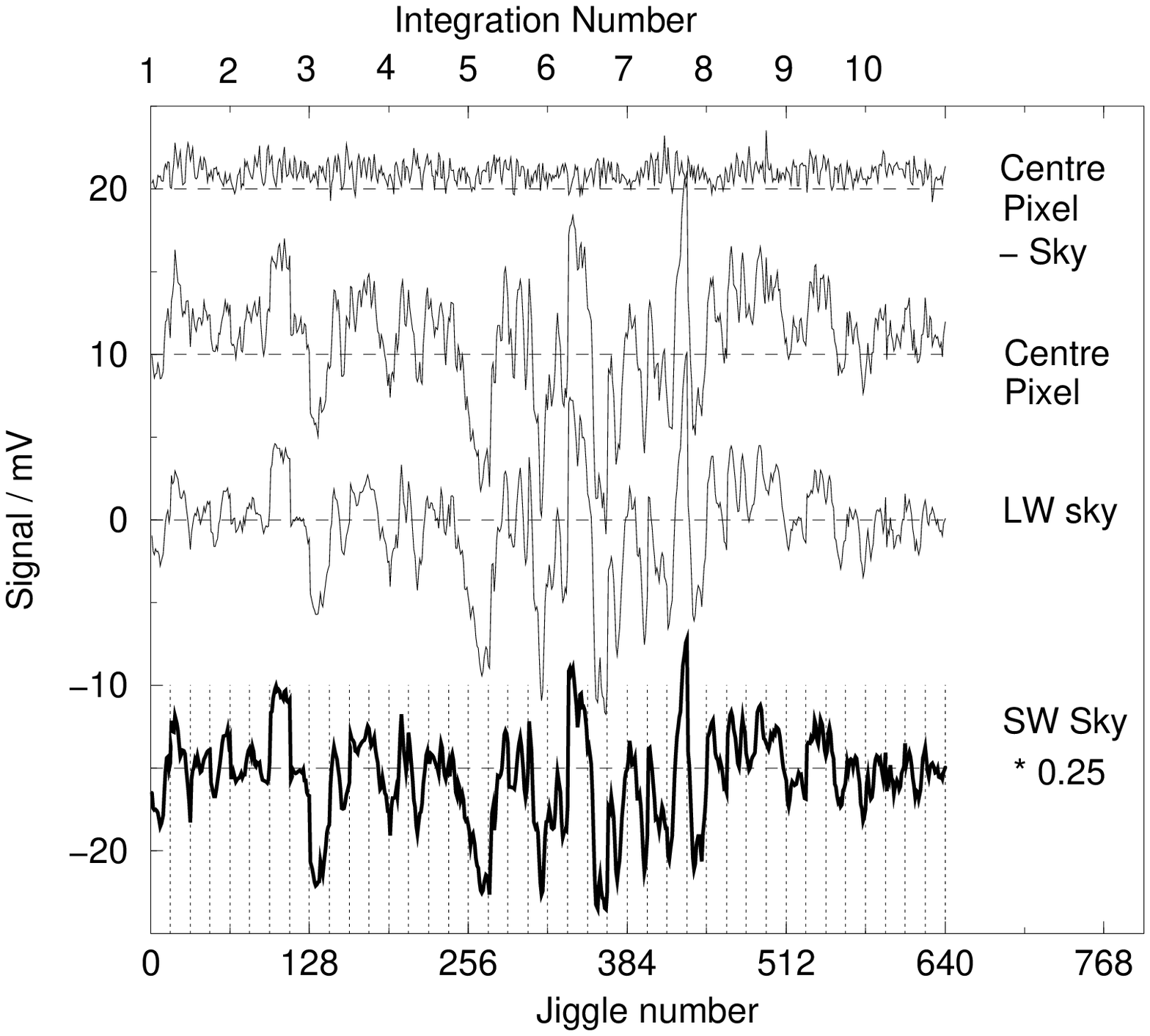}
\caption{(a) Data processed without sky removal. (b) Image after sky
contribution is removed from each jiggle position.  The contour levels are as
used in (a). (c) Plot showing data for the central pixel after sky removal and
before sky removal along with the sky signal measured for the long and short
wave arrays (lower two traces). Note that the data from the short-wave array
has been reduced by a factor of 4 to bring it onto the same scale as the
long-wave data.}
\label{fig:jigex}
\end{center}
\end{figure}

Fig.\ \ref{fig:jigex} shows a particularly good example of extreme sky
variability with Fig.\ \ref{fig:jigex}a showing the image before sky removal
and Fig.\ \ref{fig:jigex}b the image after sky removal.  The time sequence for
this observation is shown in Fig.\ \ref{fig:jigex}c: the short-wave and
long-wave sky are clearly correlated with variations in the central pixel. The
upper trace of Fig.\ \ref{fig:jigex} shows the central pixel after sky removal
and the signal variation repeats for each integration reflecting the fact
that a 64 point jiggle pattern jiggles on and off the source.

These data were taken mid-morning and the lower trace of Fig. \ref{fig:jigex}c
shows that the large scale variations occur every 16 seconds reflecting the
time taken for individual nods. Clearly, conditions were so unstable that a
shorter nod period was desirable although it was possible to remove this
effect later.

\subsection{Removing the effect of sky variability from SCUBA scan maps}

The observing technique for scan mapping is to raster the SCUBA arrays
over the map area, using the secondary chopper to measure the difference 
signal between points a short distance apart. Chopping removes the DC 
offset due to sky emission and diminishes the effect of sky variability 
on the signal. Unfortunately, it also results in a map that has the source 
profile convolved with the chop.

To restore the source profile we must deconvolve the chop from the measured
map. The problems associated with this step can best be appreciated by
considering the Fourier transform (FT) of the chop function, which is a sine
wave with zeroes at the origin and at harmonics of the inverse chop
throw. Deconvolving the chop function is equivalent to dividing the FT of the
measured map by the FT of the chop and then transforming back to image
space. Clearly, problems arise at spatial frequencies where the sine wave of
the chop FT has a low value or is zero. Noise at these frequencies is blown up
and significantly reduces the signal-to-noise of the restored map\cite{EKH79}.

A few years ago Emerson\cite{E95} proposed a method by which this problem
could be reduced and conducted tests on model data to demonstrate the
improvement. His method requires the taking of 4 maps instead of 1; 2 maps
chopping vertically with 2 different throws, and 2 chopping horizontally. The
chop amplitudes are chosen so that, except at the origin, the zeroes in the FT
of one do not coincide with the zeroes in the FT of the other up to the
spatial frequency limit of the telescope beam. The 4 maps obtained are then
Fourier transformed and coadded, each weighted in frequency space according to
the sensitivity of its chop direction and throw. The coadd is transformed back
to give the finished image.

\begin{figure}
\begin{center}
(a) \epsfig{width=3.0in,file=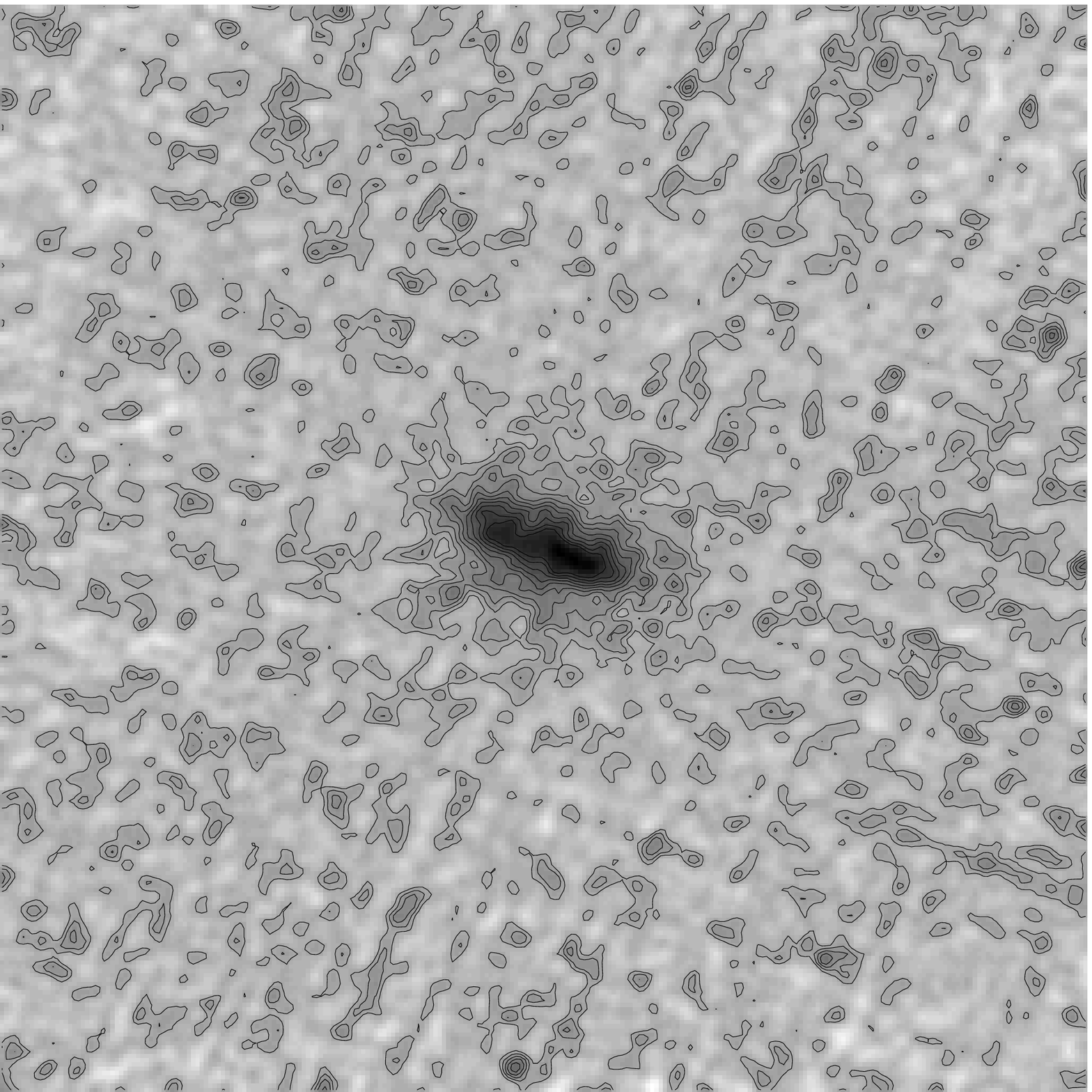}
(b) \epsfig{width=3.0in,file=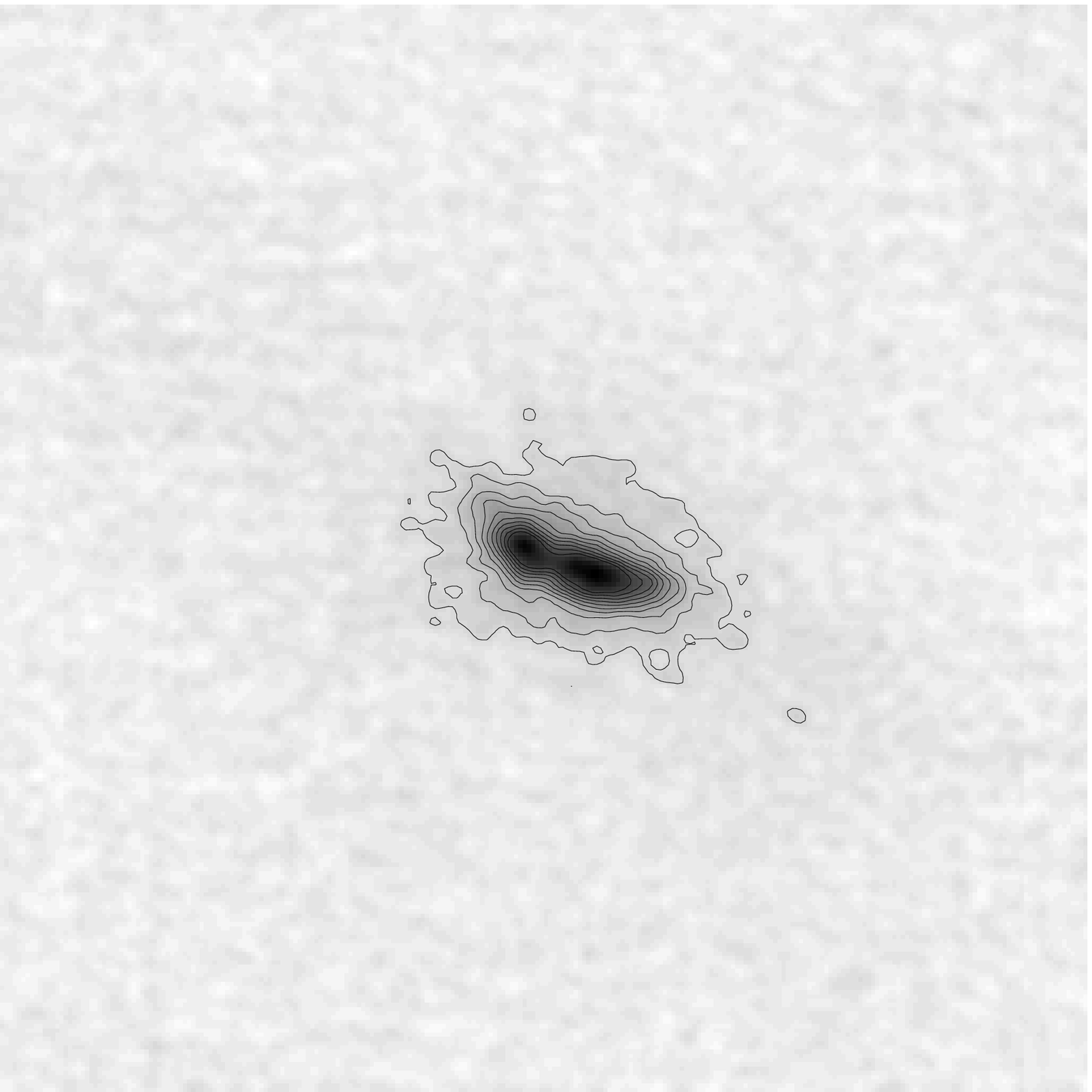}
\caption{(a) This is a 256 arcsec by 256 arcsec map of M82 taken using the
classic scan mapping technique\cite{EKH79} but with the addition of data
from a second scanning angle. (b) This is a map of the same
region with the same integration time, but taken and reduced according to the
method described by Emerson\cite{E95}. Identical contour levels have been
used.}
\label{fig:scanmaps}
\end{center}
\end{figure}

This method has been used for SCUBA with very encouraging results. The final
images are greatly improved cosmetically, are much less affected by spikes in
the raw data and, most importantly, display a significant increase in
signal-to-noise. An example of the improvement can be seen by comparing 2 maps
of a large area around the nucleus of M82, the first taken with the old method
(Fig.\ \ref{fig:scanmaps}a), the second with the new (Fig.\
\ref{fig:scanmaps}b), both with the same total integration time. In this case
the signal-to-noise of the new map is 2 times better than the old.

\begin{figure}
\begin{center}
(a) \epsfig{width=3.0in,file=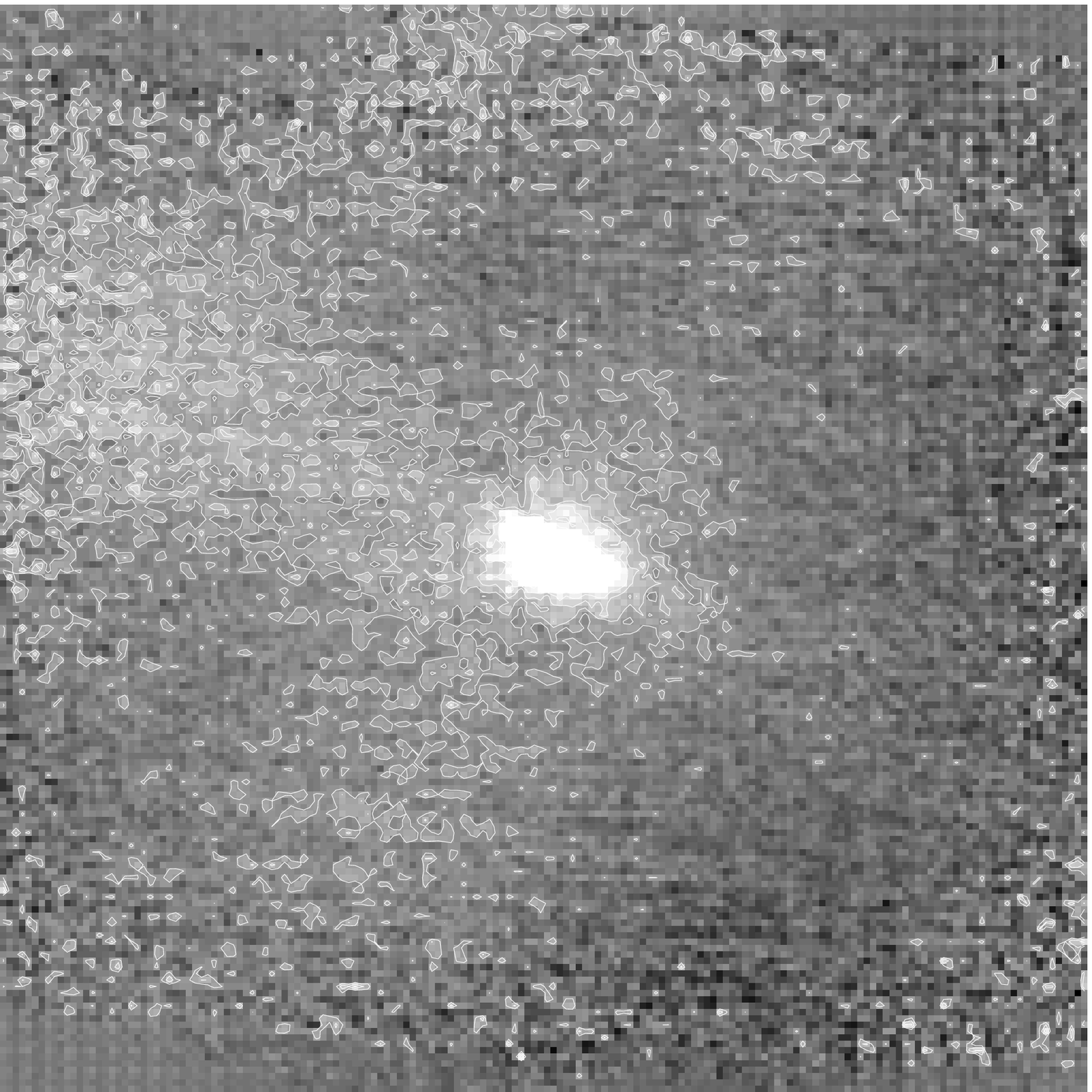}
(b) \epsfig{width=3.0in,file=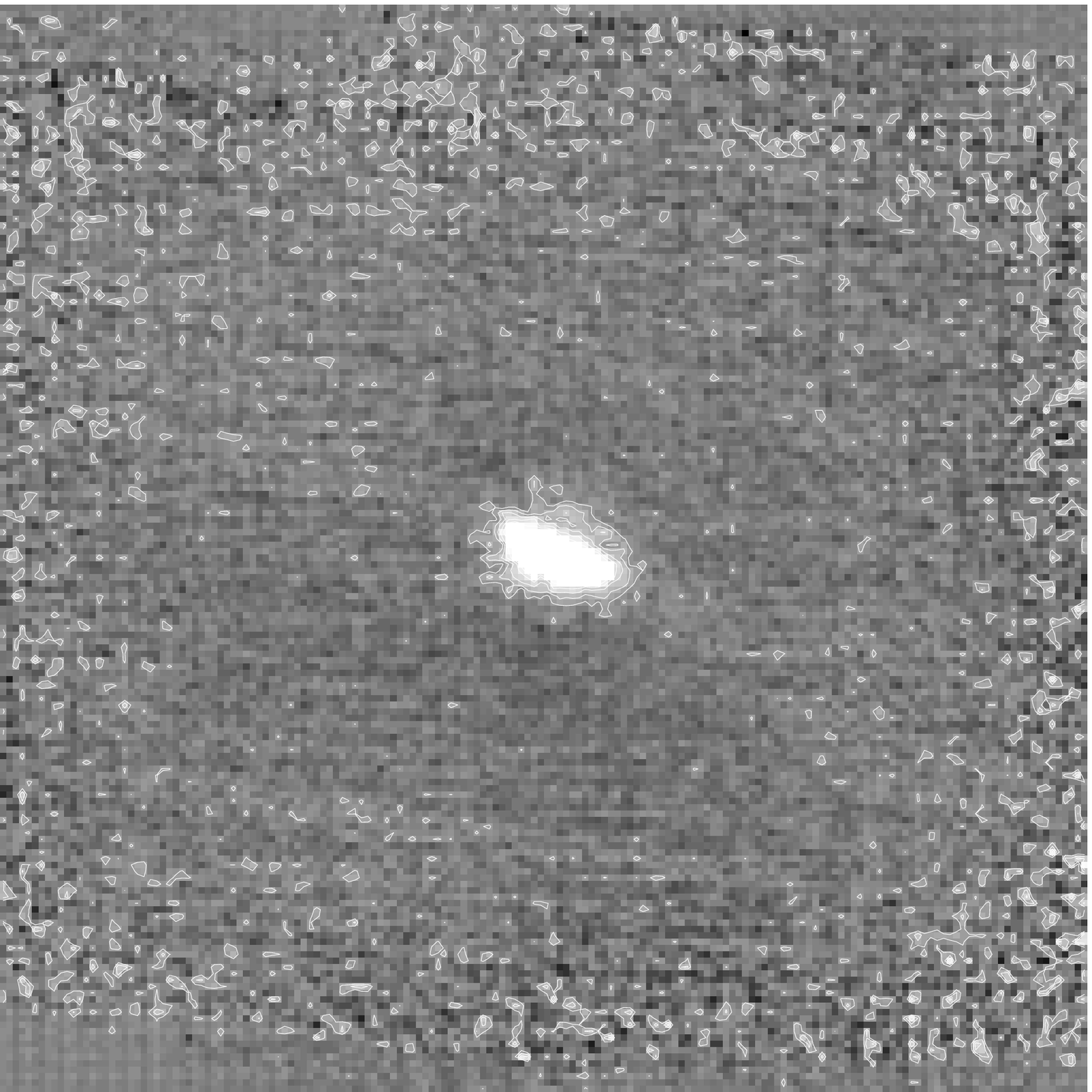}
\caption{(a) Low intensity fluctuations in the map of M82 reduced without `sky 
removal'. (b) Map of M82 after `sky removal', displayed on the same intensity
scale as (a). The contour levels are the same for each image. These images
cover a region of 500 arcsec by 500 arcsec.}
\label{fig:before_after}
\end{center}
\end{figure}

There is, however, one problem evident in the new M82 map. Strong emission
from the source is known to be restricted to a small area around the nucleus;
the rest of the map should be empty. Instead we see small amplitude, large
scale fluctuations around the zero level; to make this clearer the map is
shown again in Fig.\ \ref{fig:before_after}a with a colour table biased towards low intensity.

This residual noise is occurring at very low spatial frequencies, where 
all 4 chopped maps of the new method have low sensitivity. Even the new
method of map restoration must inflate the signal at these frequencies
to recover the source profile, along with any noise present. Unfortunately, 
it is at just such low frequencies that sky variability is likely to
contribute most noise.

\begin{figure}
\begin{center}
(a)\epsfig{file=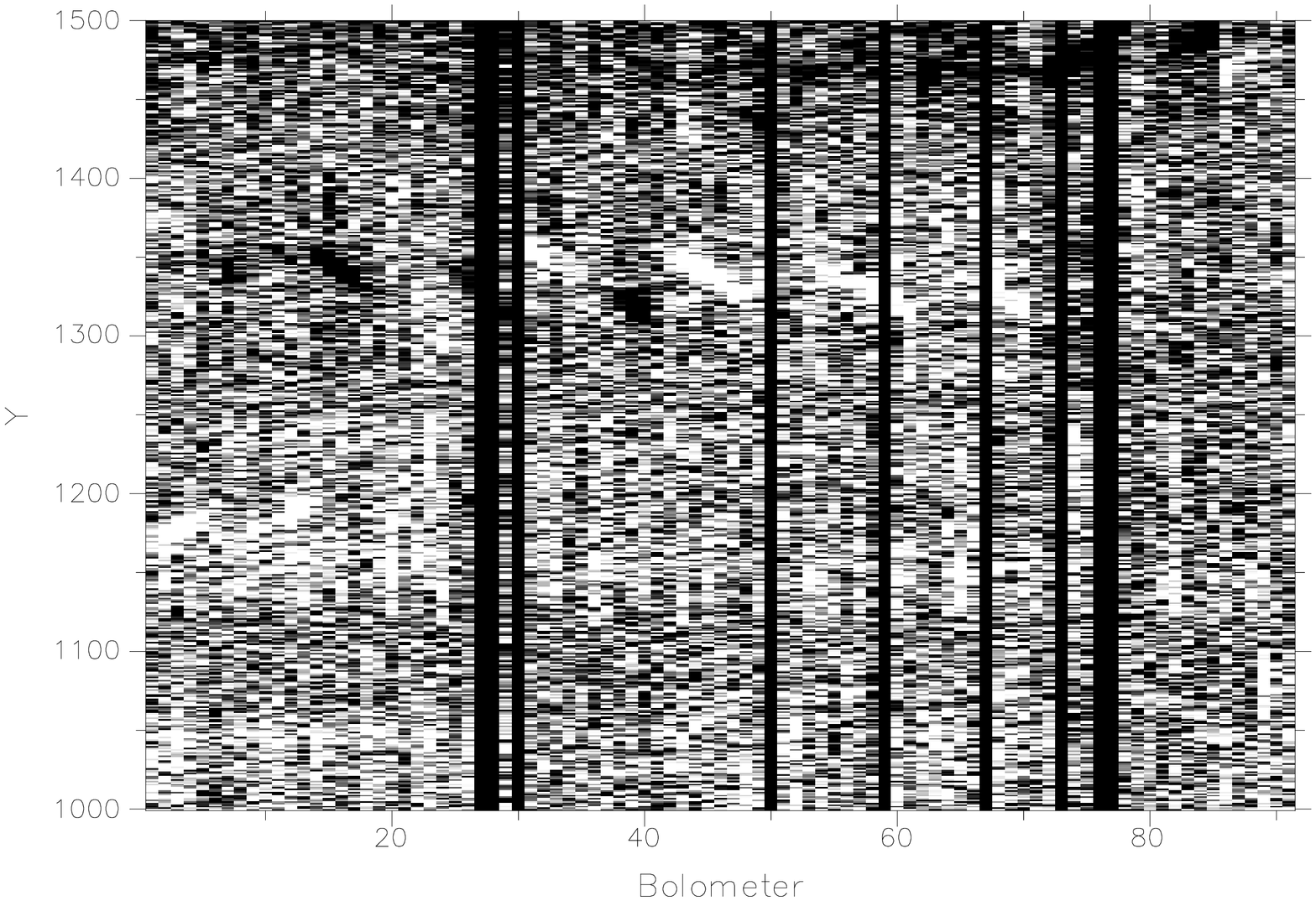,angle=-90,width=3.0in}
(b)\epsfig{file=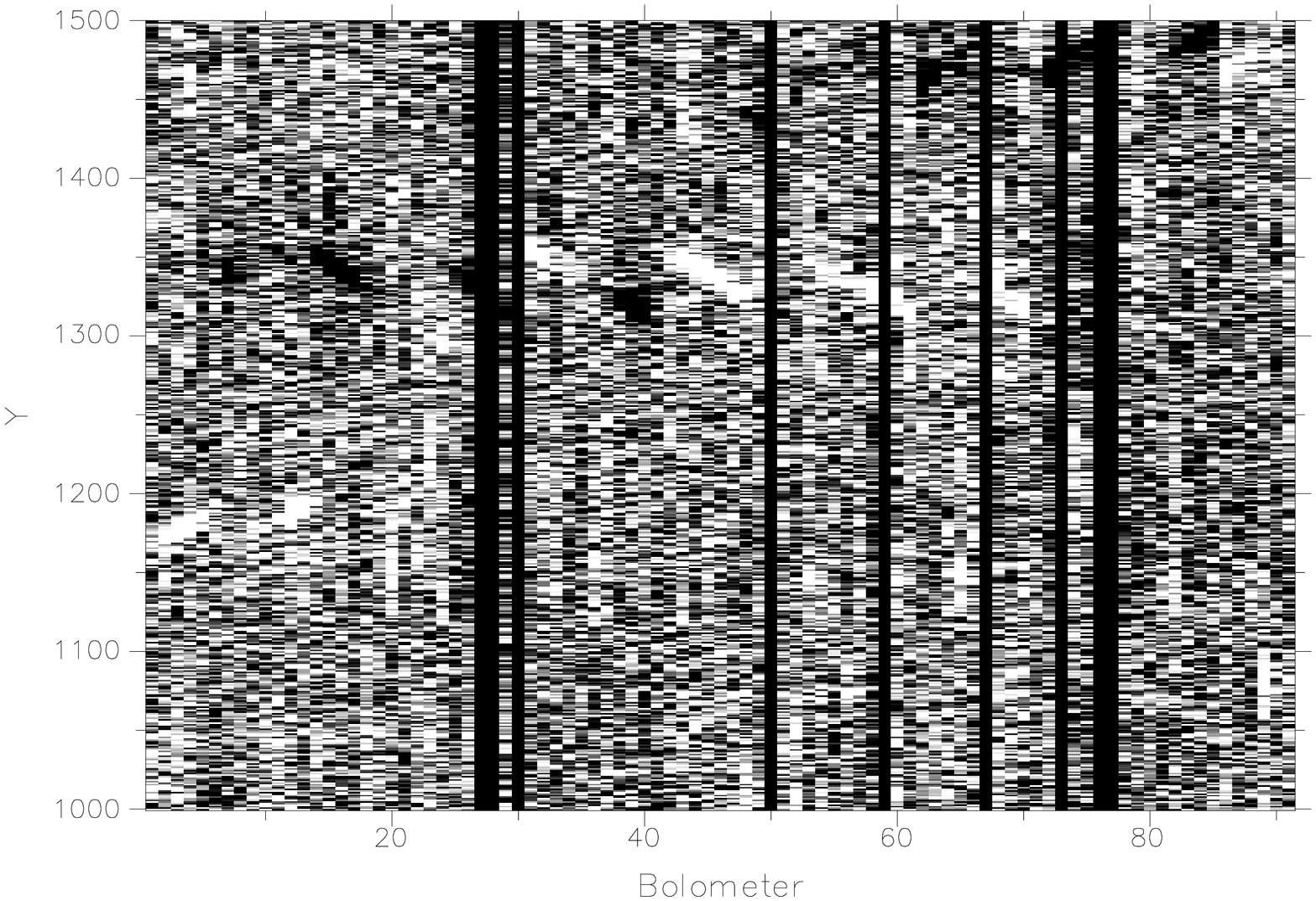,angle=-90,width=3.0in}
\caption{(a) A portion of the raw data from one of the observations of
M82. Bolometer number increases along the x-axis and measurement number (time)
up the y-axis.  The measurements were made at approximately 8~Hz. (b) The same
data after sky removal. Note that the faint banding visible across the
bolometers in (a) (e.g.\ pixels 1050--1150  and pixels 1450--1500) has
disappeared. }
\label{fig:section}
\end{center}
\end{figure}

A small section of the raw data from one of the M82 observations is shown in
Fig.\ \ref{fig:section}a. In this we can clearly see small areas of high or
low emission where bolometers have passed over the source. However, also
visible are faint light and dark bands measured across all the bolometers
simultaneously and with time-scales of order 1--20 seconds, characteristics both
of variable sky emission.
 
The main problem with removing sky variations from scan map data 
is that it is difficult to disentangle signals due to the sky from those
due to passage over the source. In general, we would have to exploit the
fact that source signals are correlated in position whereas sky 
variations correlate in time. However, our M82 dataset is a map 
covering a large area with a compact source at the centre so, in this case,
we can cheat. It is possible to effectively remove the source from the raw 
data by simply ignoring all data points with a signal further from zero than 
a given threshold.

After source removal, a `variable sky' dataset was calculated from the 
raw data by smoothing the time series signal from each bolometer with a 
box function 6 samples wide, then at each measurement time averaging the 
signal across all bolometers in the array with good noise behaviour. 
The variable sky was then subtracted from the original raw data resulting 
in a corrected dataset of which the same small section as before is plotted in
Fig.\ \ref{fig:section}b. It is evident that our crude efforts have been
successful; the banding has disappeared in the corrected data and no new
features have been added due to improper removal of source emission.

\begin{figure}
\begin{center}
\epsfig{width=5.0in,file=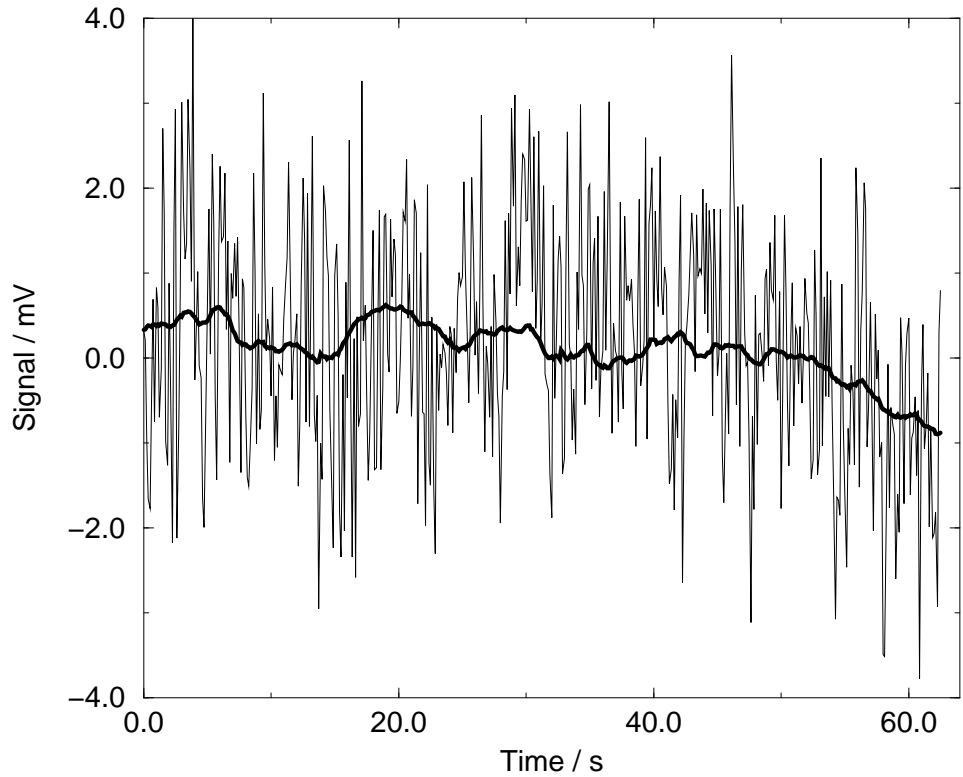}
\caption{The raw time-series data for a single bolometer with, superimposed,
the derived signal due to sky variation.}
\label{fig:scantime}
\end{center}
\end{figure}

As an example of the strength of the sky variations compared to the raw signal
from a single bolometer, the data for one bolometer over a short time-span are
shown in Fig.\ \ref{fig:scantime}. As can be seen, the sky variations are well
below the noise level for the bolometer. Smoothing in time and averaging
over the array are definitely required to measure the effect.

The method described above was used to remove the sky variations from all the
M82 datasets and these were then reduced as before to yield the result shown
in figure Fig.\ \ref{fig:before_after}b. Here we can see that sky removal has
indeed had the desired effect; the low frequency ripple on the background has
all but disappeared.

And that completes the description of our work so far. Clearly, the method of
sky removal outlined above is unsatisfactory for any but compact sources but
it does demonstrate that further effort is very worthwhile. The next stage
will be to develop a scheme by which extended sources of arbitrary shape can
be removed from the raw data. One obvious method to try will be to calculate
back from a reduced map the signal that should have been received at each
measured position in the raw data, then subtract it from the actual raw
data. Hopefully, this will be sufficiently accurate to leave the variable sky
signal behind.

\section{Conclusions}

In this work we have shown that noise due to sky variability is evident in
SCUBA data and that it is correlated across the array as well as at different
wavelengths. This raises the prospect of using sky signal measured on one
array to remove the sky signal measured on another.  Currently the biggest
issue facing us is that of sky removal for scan mapping; our preliminary
investigation has shown that we can detect and remove this sky variability,
albeit from compact sources.

%%%%%%%%%%%%%%%%%%%%%%%%%%%%%%%%%%%%%%%%%%%%%%%%%%%%%%%%%%%%%

\acknowledgments     %>>>> equivalent to \section*{ACKNOWLEDGMENTS}       

The James Clerk Maxwell Telescope is operated by The Joint Astronomy Centre on
behalf of the Particle Physics and Astronomy Research Council of the United
Kingdom, the Netherlands Organisation for Scientific Research, and the
National Research Council of Canada.

%%%%%%%%%%%%%%%%%%%%%%%%%%%%%%%%%%%%%%%%%%%%%%%%%%%%%%%%%%%%%

%%%%% References %%%%%


\begin{thebibliography}{1}

\bibitem{DRAC95}
W.~D. Duncan, E.~I. .Robson, P.~A.~R. Ade, and S.~E. Church, ``Measurements of
  submillimetre emission noise from {M}auna {K}ea,'' in {\em Multi-feed Systems
  for Radio Telescopes},  D.~T. Emerson and J.~M. Payne, eds., {\em ASP
  Conference Series} {\bf 75}, pp.~295--308, 1995.

\bibitem{HGL98}
W.~S. Holland, W.~K. Gear, J.~F. Lightfoot, T.~Jenness, E.~I. Robson, C.~R.
  Cunningham, and K.~Laidlaw, ``{SCUBA}: A submillimetre camera operating on
  the {J}ames {C}lerk {M}axwell {T}elescope,'' in {\em Advanced Technology MMW,
  Radio and Terahertz Telescopes},  {\em Proc. SPIE} {\bf this volume}, 1998.

\bibitem{CGD94}
C.~R. Cunningham, W.~K. Gear, W.~D. Duncan, P.~R. Hastings, and W.~S. Holland,
  ``{SCUBA}: The {S}ubmillimeter {C}ommon-{U}ser {B}olometer {A}rray for the
  {J}ames {C}lerk {M}axwell {T}elescope,'' in {\em Instrumentation in Astronomy
  VIII},  D.~L. Crawford and E.~R. Craine, eds., {\em Proc. SPIE} {\bf 2198},
  pp.~638--649, 1994.

\bibitem{GC95}
W.~K. Gear and C.~R. Cunningham, ``{SCUBA}: A camera for the {J}ames {C}lerk
  {M}axwell {T}elescope,'' in {\em Multi-feed Systems for Radio Telescopes},
  D.~T. Emerson and J.~M. Payne, eds., {\em ASP Conference Series} {\bf 75},
  pp.~215--221, 1995.

\bibitem{LDG95}
J.~F. Lightfoot, W.~D. Duncan, W.~K. Gear, B.~D. Kelly, and I.~A. Smith,
  ``Observing strategies for {SCUBA},'' in {\em Multi-feed Systems for Radio
  Telescopes},  D.~T. Emerson and J.~M. Payne, eds., {\em ASP Conference
  Series} {\bf 75}, pp.~327--334, 1995.

\bibitem{CLH93}
S.~E. Church, A.~N. Lasenby, and R.~E. Hills, ``An upper limit on the finescale
  anisotropy of the cosmic background radiation at 800-microns,'' {\em Mon.\
  Not.\ R.\ Astron.\ Soc.} {\bf 261}, pp.~705--717, 1993.

\bibitem{EKH79}
D.~T. Emerson, U.~Klein, and C.~G.~T. Haslam, ``A multiple beam technique for
  overcoming atmospheric limitations to single-dish observations of extended
  radio sources,'' {\em Astron.\ Astrophys.} {\bf 76}, pp.~92--105, 1979.

\bibitem{E95}
D.~T. Emerson, ``Approaches to multi-beam data analysis,'' in {\em Multi-feed
  Systems for Radio Telescopes},  D.~T. Emerson and J.~M. Payne, eds., {\em ASP
  Conference Series} {\bf 75}, pp.~309--317, 1995.

\end{thebibliography}
  \end{document}